\documentclass{nature}

\usepackage{graphicx}
\usepackage{amssymb}
\usepackage{amsmath}
\usepackage{mathrsfs}
\usepackage{multirow}
\usepackage{dcolumn}

\title{Resonances of unbound quantum many-body systems of nuclei}

\author{B. S. Hu, Q. Wu, Z. H. Sun \& F. R. Xu \thanks{frxu@pku.edu.cn}}

\begin{document}

\maketitle

\begin{affiliations}
\item[] School of Physics, and State Key Laboratory of
Nuclear Physics and Technology, Peking University, Beijing 100871,
China.
\end{affiliations}

\begin{abstract}
Resonance is a general phenomenon which can happen in classic or quantum systems. An unbound many-body quantum system can undergo a self-resonant process. It has long been a challenge how to describe unbound many-body quantum systems in resonances. In this paper, we develop a novel first-principles method that is capable of describing resonant quantum systems. We exploit, for the first time, the advanced in-medium similarity renormalization group (IMSRG) in the complex-energy Gamow-Berggren representation with resonance and non-resonant continuum. The {\it ab initio} Gamow IMSRG has broad applications, such as, to the electromagnetic-interaction systems of atoms, molecules or quantum dots, and strong-interaction atomic nuclei. In the present work, we apply the method to loosely bound or unbound nuclear systems. Carbon and oxygen isotopes have been investigated with an optimized chiral effective field theory potential. The resonant states observed in neutron-rich $^{22}$O and $^{24}$O are well reproduced. The loose halo structure of the Borromean nucleus $^{22}$C is clearly seen by the density calculation, in which the continuum $s$-waves play a crucial role. Further, we predict low-lying resonant excited states in $^{22}$C. This method provides rigorous and tractable theoretical calculations for weakly-bound or unbound open quantum systems.
\end{abstract}

\section*{Introduction}

$\qquad$ Open quantum systems (OQS) \cite{Feynman:1963fq,oqs} are ubiquitous in the nature, and their properties are profoundly affected by the environment, e.g., the scattering continuum and decay channels in atomic nuclei \cite{PhysRevLett.89.042501,PhysRevLett.89.042502}. 
OQS's have been probed in various fields, such as nuclear physics, atomic and molecular physics, quantum optics, quantum information and quantum cosmology.
Thanks to advanced radioactive beam facilities \cite{Jonson2016,nphy3165},
loosely-bound or unbound nuclei with extreme neutron-to-proton ratios have been explored in an unprecedented way.
These nuclei belong to the category of OQS in which
the coupling to the particle continuum may change the behavior of the system.
Many novel phenomena have been observed or predicted in nuclear OQS's,
such as halo \cite{PhysRevLett.55.2676,nature431}, genuine intrinsic resonance \cite{PhysRevLett.100.152502,PhysRevLett.116.102503}, shell evolution \cite{PhysRevLett.87.082502,PhysRevLett.108.242501} and new collective modes \cite{PhysRevLett.114.192502,PhysRevC.80.064311}.

Nuclear theory is pursuing {\it ab initio} calculations which are based on realistic nuclear forces \cite{PhysRevC.51.38,PhysRevC.63.024001,PhysRevC.68.041001,Machleidt20111} with rigorous many-body methods. Recently, a powerful theoretical method called similarity renormalization group (SRG)
was proposed to study light-front quantum field theory \cite{PhysRevD.48.5863} and condensed matter systems \cite{AnnPhys.506.77}.
Later, the SRG method was developed in many-body configuration space for nuclear systems,
named in-medium SRG (IMSRG) \cite{PhysRevLett.106.222502,HERGERT2016165}.
The IMSRG calculations can directly give the properties of the ground states of closed-shell nuclei,
and also be used to derive a non-perturbative effective valence-shell Hamiltonian \cite{PhysRevC.85.061304} for open-shell nuclei or excited states.
More recent IMSRG developments include multi-reference IMSRG \cite{PhysRevLett.110.242501},
IMSRG using an ensemble reference \cite{PhysRevLett.118.032502},
equation-of-motion IMSRG (EOM-IMSRG) \cite{PhysRevC.95.044304,PhysRevC.96.034324},
merging IMSRG and no-core shell model \cite{PhysRevLett.118.152503}.
The IMSRG has become a powerful and predictive {\it ab initio} method.
However, all the existing IMSRG calculations are performed in the harmonic oscillator (HO) or real-energy Hartree-Fock (HF) basis.
The real-energy bases are always bound and localized,
and hence isolated from the environment of unbound scattering states.
It is lacking to include the continuum effect in IMSRG.

In this paper, we develop IMSRG in the complex-energy space named the Beggren basis \cite{BERGGREN1968265} in which bound, resonant and non-resonant continuum states are treated on equal footing \cite{LIOTTA19961}.
The unitary IMSRG flow equation with a Hermitian Hamiltonian in the real-energy Hilbert space
is changed into an orthogonal transformation in the complex-symmetric rigged Hilbert space.
It can deal with the well-known outgoing Gamow resonances of particles. 
We call this method as Gamow IMSRG, and have applied it to carbon and oxygen isotopes.
The recent experiments \cite{PhysRevLett.104.062701,PhysRevC.86.054604,PhysRevLett.109.202503,TOGANO2016412} highlight
that $^{22}$C is the heaviest Borromean halo nucleus observed.
The experimental root-mean-squared matter radius of $^{22}$C was deduced to be $3.44\pm0.08$ fm \cite{TOGANO2016412}.
The coupling to continuum should play a role in producing the extended density distribution.
No information has been known experimentally about $^{22}$C excited states
which can provide further understanding of halo structure.
Using the Gamow IMSRG,
we give a continuum-coupled {\it ab-initio} calculation of the halo $^{22}$C for the first time.

\section*{Results}

$\qquad$ We have developed an {\it ab-initio} Gamow in-medium similarity renormalization group (Gamow IMSRG)
which includes the continuum degree of freedom via the complex-energy Gamow-Berggren Hartree-Fock (HF) space.
In the Gamow HF basis, by using the Gamow IMSRG the continuum-coupled Hamiltonian of a $A$-nucleon system is decoupled first with the ground-state configuration, giving the ground-state property of nucleus. Then, with the decoupled IMSRG Hamiltonian, we perform the equation of motion (EOM) calculation to obtain excited states, which we call Gamow EOM-IMSRG.
The method provides a unified description of bound, resonant and continuum states.
We hope that the Gamow EOM-IMSRG can provide a reliable {\it ab-initio} calculation for loosely-bound or unbound quantum systems.
In the present work, we focus on weakly-bound carbon and oxygen isotopes in which the resonance phenomenon has been observed in experiment.

\begin{figure}
\centering
\setlength{\abovecaptionskip}{0pt}
\setlength{\belowcaptionskip}{0pt}
\includegraphics[scale=0.65]{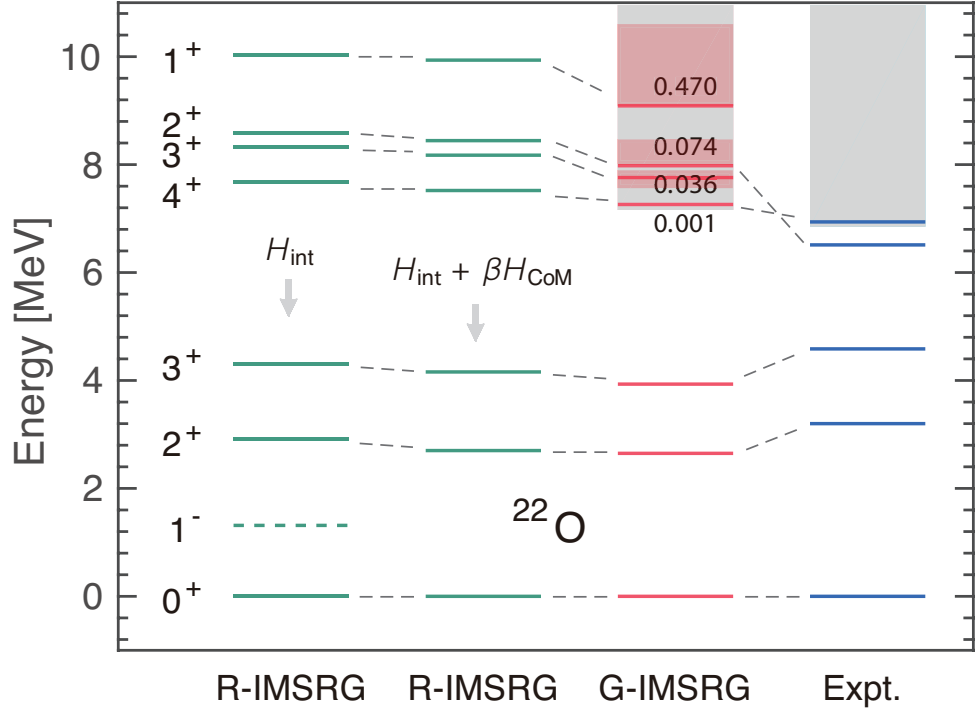}
\caption{ $^{22}$O excitation spectra calculated with the chiral NNLO$_{\rm{opt}}$ interaction.
The left two columns give the real-energy EOM-IMSRG results (indicated by R-IMSRG) with and without the Lawson center-of-mass (CoM) treatment, $\beta H_{\text{CoM}}$ taking $\beta=5.0$.
The right two columns display the Gamow EOM-IMSRG calculations (indicated by G-IMSRG) and experimental data \cite{nndc}.
The resonance states are indicated by shading, and their width (in MeV) are given by the numbers near the levels. The gray shading means the particle continuum above the scattering threshold.}
\label{o22_n2lo-opt_eom_hcm_continua}
\end{figure}
\begin{figure}
\centering
\setlength{\abovecaptionskip}{0pt}
\setlength{\belowcaptionskip}{0pt}
\includegraphics[scale=0.70]{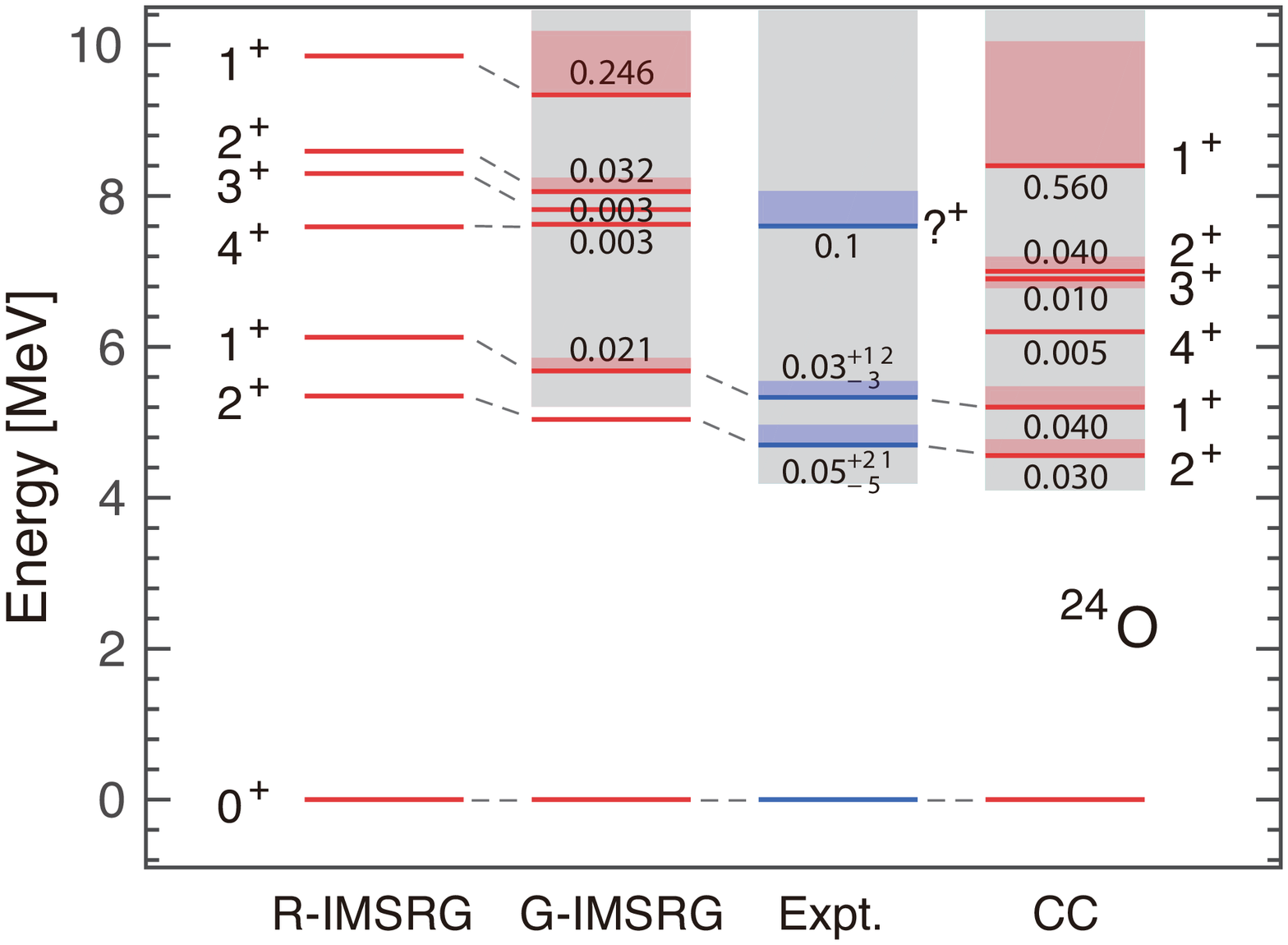}
\caption{Excited states in $^{24}$O calculated by Gamow EOM-IMSRG, compared with the complex coupled-cluster (CC) calculation \cite{PhysRevLett.108.242501}.
The data are taken from \cite{HOFFMAN200917,PhysRevC.83.031303}. The labels are similar to in Figure~\ref{o22_n2lo-opt_eom_hcm_continua}.}
\label{o24_n2lo-opt_eom_cc_continua}
\end{figure}
Figures~\ref{o22_n2lo-opt_eom_hcm_continua} and \ref{o24_n2lo-opt_eom_cc_continua} give the Gamow EOM-IMSRG calculations for $^{22}$O and $^{24}$O with the state-of-the-art chiral NNLO$_{\rm opt}$ interaction \cite{PhysRevLett.110.192502}. 
The calculation reproduces the experimental observations of the excitation energies and their resonances.
In Ref.~\cite{PhysRevLett.108.242501}, Hagen {\it et al.} performed a complex coupled-cluster (CC) calculation for $^{24}$O.
The calculation includes a schematic three-nucleon force.
We see that the two types of calculations by Gamow EOM-IMSRG and complex CC give similar results.
The large excitation energies of the first $2^+$ states in $^{22}$O and $^{24}$O support
the shell closures at $N$=14 and 16 in the oxygen chain.
In $^{24}$O, the present calculations give three resonant states with $J^{\pi}=2^{+}, 3^{+}$ and $4^{+}$,
near the not-yet-clear experimental levels at $\sim 7.6$ MeV \cite{PhysRevC.83.031303}.
This prediction is consistent with the complex CC calculation \cite{PhysRevLett.108.242501}.
By comparing the results with (G-IMSRG) and without (R-IMSRG) the continnum coupling in figures~\ref{o22_n2lo-opt_eom_hcm_continua} and \ref{o24_n2lo-opt_eom_cc_continua}, we can clearly see the continuum effect in the $^{22,24}$O.

\begin{figure}
\centering
\setlength{\abovecaptionskip}{0pt}
\setlength{\belowcaptionskip}{0pt}
\includegraphics[scale=0.70]{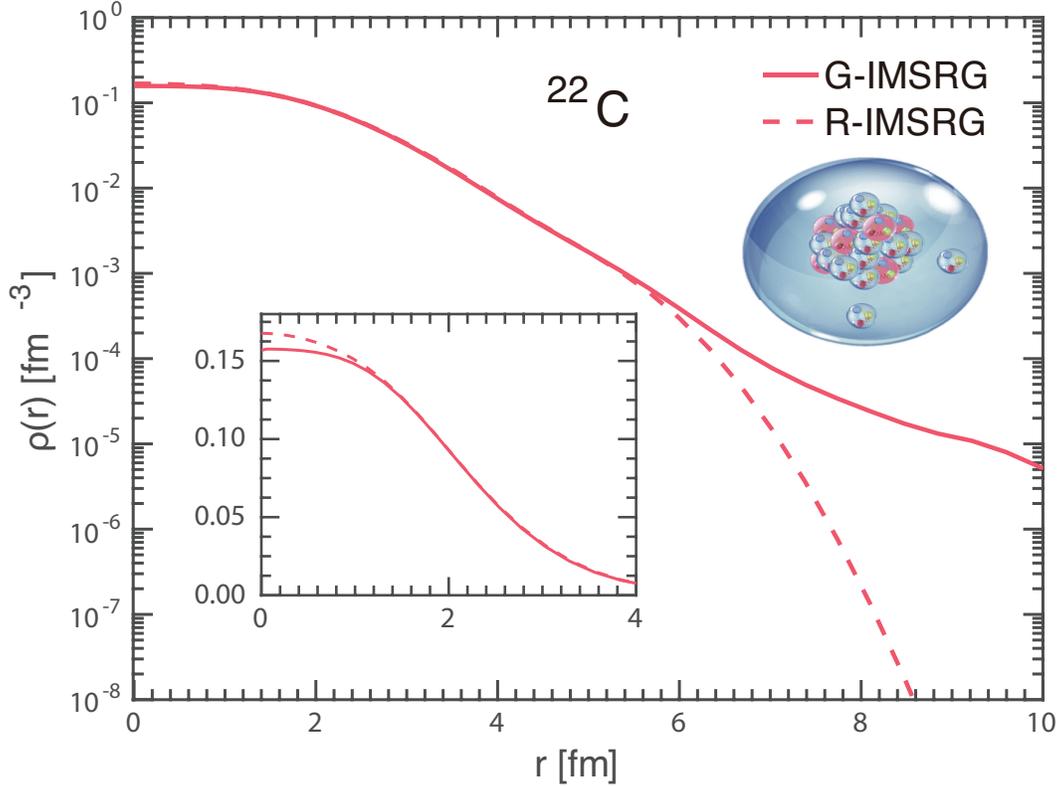}
\caption{Calculated $^{22}$C ground-state density displayed in the logarithm scale.
The ``R-IMSRG" indicates the real-energy IMSRG calculation,
while ``G-IMSRG" is the complex Gamow IMSRG calculation.
The inset details the density in the central region of the nucleus with the standard scale.}
\label{density}
\end{figure}
With the successes of the $NNLO_{\text opt}$ Gamow IMSRG calculations in $^{22,24}$O,
we have also extended the current calculation to the Borromean halo $^{22}$C which is a challenging nucleus for many theoretical calculations \cite{KUO199615,Sun:2018ekv}.
The present Gamow Hartree-Fock (GHF) calculation gives that the neutron $\nu 1s_{1/2}$ orbital is weakly bound.  The orbital is almost fully occupied in the Gamow IMSRG ground state of $^{22}$C.
The two-neutron configuration $[\nu 1s_{1/2}]^{2}$ is responsible for halo formation \cite{PhysRevLett.104.062701,PhysRevC.86.054604,PhysRevLett.109.202503,TOGANO2016412}.
We have calculated the $^{22}$C ground-state density with an effective density operator that can be derived within the real-energy IMSRG or the complex-energy Gamow IMSRG, shown in Figure~\ref{density}.
It is seen that the Gamow IMSRG calculation gives a long tail in the density, indicating a halo structure.

To see the effect from the continuum, we have analyzed the $s$ channel. Two types of the Gamow IMSRG calculations have been performed, 
with discrete HF $s$-waves which are obtained in the real-energy HF calculation or continuum GHF $s$-waves that are obtained in the GHF calculation.
In both calculations, the neutron $d_{3/2}$ channel remains in the Gamow HF basis.
The $s$-waves play the key role in the halo structure.
The Gamow IMSRG calculation with adopting discrete HF $s$-waves (instead of continuum Gamow $s$-waves) gives a matter radius of 2.79 fm for the $^{22}$C ground state,
while the calculation with taking the continuum $s$-waves gives a larger radius of 3.06 fm.
The experimental estimated  matter radius is $5.4\pm0.9$ fm in an earlier work \cite{PhysRevLett.104.062701}
and it is $3.44\pm0.08$ fm in the later work \cite{TOGANO2016412}.
The NNLO$_{\text{opt}}$ interaction itself underestimates the radii of carbon isotopes \cite{PhysRevLett.117.102501}.

There have been no experimental data available for excited states in $^{22}$C.
Table~\ref{Ex_c22} gives the Gamow EOM-IMSRG prediction for possible low-lying states.
The first $2^{+}$ state is bound, and its energy is lower than other calculations \cite{PhysRevC.81.064303,PhysRevLett.113.142502}.
We find that the $2_1^+$ state is dominated by $1p$-$1h$ excitation from the proton $\pi 0p_{3/2}$ hole to $\pi 0p_{1/2}$ particle orbitals.
The excitation of the proton orbitals gives lower energy.
In addition, we predict a superposition of narrow resonances with $J^{\pi}=1^{+}$, $2^+$, $3^+$ and $4^{+}$
around the energy position of 3.5 MeV and a width of $\sim 0.25$ MeV.
These resonances are dominated by $1p$-$1h$ excitations from the neutron $\nu 0d_{5/2}$ hole to $\nu 0d_{3/2}$ particle orbitals.
\begin{table}
\centering
\caption{
\label{Ex_c22}
Excited states for $^{22}$C predicted by the Gamow EOM-IMSRG with the chiral NNLO$_{\rm{opt}}$ interaction.
Energies and widths are in MeV.
}
\begin{tabular}{ccccccccc p{cm}}
\hline
\hline
$J^{\pi}$& $2^{+}_{1}$& $4^{+}_{1}$& $3^{+}_{1}$& $1^{+}_{1}$& $2^{+}_{2}$& $2^{+}_{3}$& $1^{+}_{2}$\\
\hline
$E_{\text{IMSRG}}        $& 1.05 & 3.26 & 3.36 & 3.62 & 3.93 & 4.94 & 5.27\\
\hline
$\varGamma_{\text{IMSRG}}$& 0.000& 0.245& 0.262& 0.135& 0.266& 0.597& 0.604\\
\hline
\hline
\end{tabular}
\end{table}

\section*{Discussion}

$\qquad$ The intrinsic Hamiltonian of a $A$-nucleon system reads
\begin{eqnarray}
\text{\^{H}}=
\displaystyle\sum_{i=1}^{A} \left(1-\dfrac{1}{A}\right) \dfrac{\vec{p}_{i}}{2m} +
\displaystyle\sum_{i<j=1}^{A}   \left(\hat{V}_{ij}-\dfrac{\vec{p}_{i}\cdot\vec{p}_{j} }{mA} \right),
\label{Hamiltonian}
\end{eqnarray}
where $\hat{V}=\hat{V}_{NN}+\hat{V}_{\text{Coul}}$ with $\hat{V}_{NN}$ being the nucleon-nucleon ($NN$) interaction
and $\hat{V}_{\text{Coul}}$ for the Coulomb potential.
In the present work, we use the optimized chiral $NN$ interaction NNLO$_{\text{opt}}$ \cite{PhysRevLett.110.192502}. 
It yields $\chi^{2} \approx 1$ per degree of freedom for laboratory energies below approximately 125 MeV
in phase-shift analyses by the optimization tool of the Practical Optimization Using No Derivatives (for Squares) algorithm \cite{PhysRevLett.110.192502}.
The NNLO$_{\text{opt}}$ potential gives good descriptions of nuclear structures
including binding energies, radii, excitation spectra, dripline positions and the neutron matter equation of state,
without resorting to three-body forces \cite{PhysRevLett.110.192502}. 
In our calculations, a total number of 12 major HO shells at $\hbar\Omega= 20$ MeV are taken to expand the HF basis.
The convergence in such basis space has been discussed in our previous works \cite{1674-1137-41-10-104101,PhysRevC.94.014303}.

In Ref.~\cite{Morris}, the IMSRG approximation named MAGNUS(2*) was suggested,
in which a class of undercounted terms in the normal-ordered two-body truncation are restored
by introducing an auxiliary one-body operator with hole-hole and particle-particle excitations \cite{jcp141,PhysRevC.95.044304}.
MAGNUS(2*) includes more correlations which are from intermediate three-body forces \cite{Morris}.
In the present work, we use the MAGNUS(2*) method
to derive the Gamow IMSRG Hamiltonian which is decoupled with the ground state of the nucleus.
With the Gamow IMSRG Hamiltonian obtained thus, we perform the EOM calculation for the resonance and continuum many-body system.
Note that only hole-hole excitations are considered in the auxiliary one-body operator.
In Ref.~\cite{EVANGELISTA201227}, it has been shown that the contribution of particle-particle excitations is significantly smaller than the hole-hole contribution (see also the supplementary material).
On the other hand, particle-particle excitations involve a huge numbers of complex continuum states,
which induces some double counting of high-order energy contributions \cite{EVANGELISTA201227} and may bring a non-physics large imaginary component even in the ground state of a bound nucleus.
In practical calculations, for neutron-rich carbon and oxygen isotopes,
the neutron $d_{3/2}$ and $s_{1/2}$ partial waves
are treated in the resonance and continuum Berggren basis,
while other neutron channels and all proton channels are treated in the real-energy discrete HF basis.
The $0d_{3/2}$ orbit is a resonant state,
while the $1s_{1/2}$ level should have a significant contribution to the extended spatial distribution of loosely-bound or unbound resonant nuclei.
Such treatment is similar to in the no-core Gamow shell model (NCGSM) and complex coupled-cluster (CC) calculations \cite{PhysRevC.88.044318,PhysRevLett.108.242501}
where only the dominating orbits are treated in the Berggren basis.

Although Hamiltonian (\ref{Hamiltonian}) is intrinsic,
the IMSRG wave function is expressed in the laboratory coordinate.
We should consider the effect from the center-of-mass (CoM) motion which can cause spurious excitations \cite{0954-3899-16-2-013}.
In the HO basis, the CoM motion can be treated using the Lawson method by adding a multiplied Hamiltonian of the CoM motion \cite{lawson}.
In the HF basis, there is no exact method to treat the CoM effect.
However, it has been shown that, in the real-energy HF basis,
the many-body wave function can be factorized approximately into a product of an intrinsic wave function and a CoM Gaussian \cite{PhysRevLett.103.062503,HERGERT2016165},
which indicates that one can approximately use the Lawson method to treat the CoM motion.
Unfortunately, the Lawson method cannot be used in the complex-energy Berggren basis,
due to the fact that the $R^2$ matrix elements (R is the CoM position) cannot be regularized in the square non-integrable resonance and continuum states.
In the previous work \cite{SUN2017227}, however, we have discussed that the CoM effect within an intrinsic Hamiltonian is small for low-lying states.
We can estimate the CoM effect in the IMSRG calculation.
Figure~\ref{o22_n2lo-opt_eom_hcm_continua} shows the real-energy EOM-IMSRG calculations without and with the Lawson CoM term,
$\beta H_{\rm{CoM}}$
with $H_{\text{CoM}}=\dfrac{\vec{P}}{2mA}+\dfrac{1}{2}mA \tilde{\Omega}^{2}\vec{R}^{2}-\dfrac{3}{2}\hbar\tilde{\Omega}$
($\tilde\Omega$ for the CoM vibration frequency).
The $\tilde\Omega$ value can be different from the frequency of the HO basis
in which the HF equation is solved \cite{PhysRevLett.103.062503}.
We see that, in the real-energy EOM-IMSRG calculation without the CoM treatment, there is a $1^-$ state,
while the $1^-$ state does not appear in the calculation with the CoM treatment (in fact it is pushed up).
The $1^-$ state is a spurious state of the CoM excitation.
The spurious state can be easily identified and projected out in the practical calculations.
By comparing the real-energy EOM-IMSRG calculations, the CoM effect is small.
The Gamow EOM-IMSRG calculations give similar results to the real-energy EOM-IMSRG calculations for the bound states,
but predict the resonance properties of the unbound excited states.

\section*{Methods}

\subsection{{\it Gamow Hartree-Fock.$-$}}
In the Gamow calculations, how to choose a proper one-body potential to generate the resonance and continuum Berggren basis is a key step.
Usually, the Woods-Saxon potential is used \cite{PhysRevLett.89.042501,PhysRevLett.89.042502,PhysRevC.96.044307}.
However, one has to adjust the Woods-Saxon parameters to reproduce bound and unbound resonant single-particle levels
that would be lacking in experiment.
It is more self-consistent when we construct the Berggren basis using the same realistic interaction as NNLO$_{\text{opt}}$.
In the present paper, the Gamow Hartree-Fock (GHF) method \cite{PhysRevC.73.064307} has been developed for this goal.
To begin with, a HF approximation for Hamiltonian (1) is performed in the HO basis.
The HF single-particle state $|\alpha \rangle$ is expressed in the HO basis $|p\rangle$,
\begin{eqnarray}
|\alpha \rangle=\displaystyle\sum_{p}D_{p\alpha}|p\rangle.
\end{eqnarray}
The coefficients $D_{p\alpha}$ are determined by digonalizing the standard HF mean-field equation.
Then, we can obtain the HF one-body potential in the HO basis $|p\rangle$,
\begin{eqnarray}
\langle p|U|q\rangle
=\displaystyle\sum_{i=1}^{A}\displaystyle\sum_{rs}\langle pr|V|qs\rangle
D^{\ast}_{ri}D_{si},
\end{eqnarray}
where $V$ is the two-body part of Hamiltonian (1).
The HF one-body Hamiltonian in the complex-momentum (complex-$k$) space is given by
\begin{eqnarray}
\langle k|h|k'\rangle=(1-\dfrac{1}{A})\dfrac{\hbar^{2}k^{2} }{2m} \delta_{kk'}
+\displaystyle\sum_{pq}\langle p|U|q\rangle \langle k|p \rangle \langle q| k' \rangle,
\label{ghf}
\end{eqnarray}
where $\langle k| p \rangle$ is the HO basis wavefunction $|p\rangle$ expressed in the complex-$k$ space $\langle k|$.
In practical calculations, the momentum is discretized in the contour of the Berggren complex-$k$ plane \cite{SUN2017227}.
Bound, resonant and continuum HF basis states can be obtained by diagonalizing the complex-energy HF Hamiltonian (\ref{ghf}).

\subsection{{\it Gamow IMSRG.$-$}}
The philosophy behind SRG is to suppress off-diagonal matrix elements,
and drive the original Hamiltonian $H(0)=H$ towards a band- or block-diagonal form
by means of a continuous similarity transformation $U(s)$ with $U(s) \cdot U^{-1}(s)=1$
\cite{PhysRevD.48.5863,AnnPhys.506.77,PhysRevLett.106.222502,HERGERT2016165},
\begin{eqnarray}
H(s)=U(s)H(0)U^{-1}(s).
\end{eqnarray}
The Hamiltonian in the real-energy space is Hermitian, $H=H^{\dag}$,
therefore the similarity transformation  $U(s)$ can be a unitary transformation in practice,
$U(s) \cdot U^{\dag}(s)=U(s) \cdot U^{-1}(s)=1$, giving
\begin{eqnarray}
H(s)=U(s)H(0)U^{\dag}(s).
\end{eqnarray}
Differentiation yields the operator flow equation,
\begin{eqnarray}
\frac{dH(s)}{ds}=[\eta(s),H(s)],
\label{eq1:imsrg}
\end{eqnarray}
with the anti-Hermitian generator $\eta(s)$,
\begin{eqnarray}
\eta(s)=\dfrac{dU(s)}{ds}U^{\dag}(s)=-\eta^{\dag}(s).
\end{eqnarray}
Here, we extend the IMSRG to the complex Berggren basis
in which the Hamiltonian becomes complex symmetric, $H=H^{T}$.
We need a continuous orthogonal transformation, $U(s) \cdot U^{T}(s)=U(s) \cdot U^{-1}(s)=1$,
to render the $H(0)$ diagonal,
\begin{eqnarray}
H(s)=U(s)H(0)U^{T}(s).
\end{eqnarray}
Correspondingly, the generator appearing in the operator flow equation (\ref{eq1:imsrg}) becomes
\begin{eqnarray}
\label{eta}
\eta(s)=\dfrac{dU(s)}{ds}U^{T}(s)=-\eta^{T}(s).
\end{eqnarray}

We refer to the IMSRG processed in the resonance and continuum configuration space as Gamow IMSRG.
The IMSRG itself can calculate the ground state of a closed-shell nucleus
by decoupling the Hamiltonian with the closed shell \cite{PhysRevLett.106.222502}.
To calculate open-shell nuclei or excited states,
we explore the Gamow IMSRG within the equation of motion (EOM), named Gamow EOM-IMSRG.
The Magnus IMSRG formulation \cite{PhysRevC.92.034331}
and the White generator $\eta(s)$ \cite{HERGERT2016165} are adopted to decouple the Hamiltonian.
By using the White generator, Eq.~(\ref{eta}) can be satisfied easily when the Hamiltonian is extended to be complex symmetric.
With the White generator, the off-diagonal elements of the Hamiltonian are suppressed with a decay scale $(s-s_{0})$ \cite{HERGERT2016165}
whether in real or complex energy.
The general framework of real-energy EOM-IMSRG can be found in Refs. \cite{PhysRevC.95.044304,PhysRevC.96.034324}.
In the Gamow EOM-IMSRG calculations, we truncate the excitation operator at a two-particle two-hole level.

\section*{References}


\bibliography{references}

\begin{thebibliography}{10}
\expandafter\ifx\csname url\endcsname\relax
  \def\url#1{\texttt{#1}}\fi
\expandafter\ifx\csname urlprefix\endcsname\relax\def\urlprefix{URL }\fi
\providecommand{\bibinfo}[2]{#2}
\providecommand{\eprint}[2][]{\url{#2}}

\bibitem{Feynman:1963fq}
\bibinfo{author}{Feynman, R.~P.} \& \bibinfo{author}{Vernon, F.~L., Jr.}
\newblock \bibinfo{title}{{The theory of a general quantum system interacting
  with a linear dissipative system}}.
\newblock \emph{\bibinfo{journal}{Annals of Physics}}
  \textbf{\bibinfo{volume}{24}}, \bibinfo{pages}{118--173}
  (\bibinfo{year}{1963}).

\bibitem{oqs}
\bibinfo{author}{\'Angel, R.} \& \bibinfo{author}{Susana, F.~H.}
\newblock \emph{\bibinfo{title}{Open Quantum Systems}}
  (\bibinfo{publisher}{Springer-Verlag Berlin Heidelberg},
  \bibinfo{year}{2012}).
\newblock \urlprefix\url{https://www.springer.com/gp/book/9783642233531}.

\bibitem{PhysRevLett.89.042501}
\bibinfo{author}{Id~Betan, R.}, \bibinfo{author}{Liotta, R.~J.},
  \bibinfo{author}{Sandulescu, N.} \& \bibinfo{author}{Vertse, T.}
\newblock \bibinfo{title}{Two-particle resonant states in a many-body mean
  field}.
\newblock \emph{\bibinfo{journal}{Phys. Rev. Lett.}}
  \textbf{\bibinfo{volume}{89}}, \bibinfo{pages}{042501}
  (\bibinfo{year}{2002}).

\bibitem{PhysRevLett.89.042502}
\bibinfo{author}{Michel, N.}, \bibinfo{author}{Nazarewicz, W.},
  \bibinfo{author}{P\l{}oszajczak, M.} \& \bibinfo{author}{Bennaceur, K.}
\newblock \bibinfo{title}{Gamow shell model description of neutron-rich
  nuclei}.
\newblock \emph{\bibinfo{journal}{Phys. Rev. Lett.}}
  \textbf{\bibinfo{volume}{89}}, \bibinfo{pages}{042502}
  (\bibinfo{year}{2002}).

\bibitem{Jonson2016}
\bibinfo{author}{Jonson, B.}
\newblock \bibinfo{title}{What's next in nuclear physics with \text{RIB}'s}.
\newblock \emph{\bibinfo{journal}{The European Physical Journal Plus}}
  \textbf{\bibinfo{volume}{131}}, \bibinfo{pages}{20} (\bibinfo{year}{2016}).

\bibitem{nphy3165}
\bibinfo{author}{Jenkins, D.~G.}
\newblock \bibinfo{title}{Recent advances in nuclear physics through on-line
  isotope separation}.
\newblock \emph{\bibinfo{journal}{Nature Physics}}
  \textbf{\bibinfo{volume}{10}}, \bibinfo{pages}{909--913}
  (\bibinfo{year}{2014}).

\bibitem{PhysRevLett.55.2676}
\bibinfo{author}{Tanihata, I.} \emph{et~al.}
\newblock \bibinfo{title}{Measurements of interaction cross sections and
  nuclear radii in the light $p$-shell region}.
\newblock \emph{\bibinfo{journal}{Phys. Rev. Lett.}}
  \textbf{\bibinfo{volume}{55}}, \bibinfo{pages}{2676--2679}
  (\bibinfo{year}{1985}).

\bibitem{nature431}
\bibinfo{author}{Hinde, D.} \& \bibinfo{author}{Dasgupta, M.}
\newblock \bibinfo{title}{Nuclear physics: Neutron halo slips}.
\newblock \emph{\bibinfo{journal}{Nature}} \textbf{\bibinfo{volume}{431}},
  \bibinfo{pages}{748--751} (\bibinfo{year}{2004}).

\bibitem{PhysRevLett.100.152502}
\bibinfo{author}{Hoffman, C.~R.} \emph{et~al.}
\newblock \bibinfo{title}{Determination of the ${N=16}$ shell closure at the
  oxygen drip line}.
\newblock \emph{\bibinfo{journal}{Phys. Rev. Lett.}}
  \textbf{\bibinfo{volume}{100}}, \bibinfo{pages}{152502}
  (\bibinfo{year}{2008}).

\bibitem{PhysRevLett.116.102503}
\bibinfo{author}{Kondo, Y.} \emph{et~al.}
\newblock \bibinfo{title}{Nucleus $^{26}\mathrm{O}$: A barely unbound system
  beyond the drip line}.
\newblock \emph{\bibinfo{journal}{Phys. Rev. Lett.}}
  \textbf{\bibinfo{volume}{116}}, \bibinfo{pages}{102503}
  (\bibinfo{year}{2016}).

\bibitem{PhysRevLett.87.082502}
\bibinfo{author}{Otsuka, T.} \emph{et~al.}
\newblock \bibinfo{title}{Magic numbers in exotic nuclei and spin-isospin
  properties of the $\mathit{NN}$ interaction}.
\newblock \emph{\bibinfo{journal}{Phys. Rev. Lett.}}
  \textbf{\bibinfo{volume}{87}}, \bibinfo{pages}{082502}
  (\bibinfo{year}{2001}).

\bibitem{PhysRevLett.108.242501}
\bibinfo{author}{Hagen, G.}, \bibinfo{author}{Hjorth-Jensen, M.},
  \bibinfo{author}{Jansen, G.~R.}, \bibinfo{author}{Machleidt, R.} \&
  \bibinfo{author}{Papenbrock, T.}
\newblock \bibinfo{title}{Continuum effects and three-nucleon forces in
  neutron-rich oxygen isotopes}.
\newblock \emph{\bibinfo{journal}{Phys. Rev. Lett.}}
  \textbf{\bibinfo{volume}{108}}, \bibinfo{pages}{242501}
  (\bibinfo{year}{2012}).

\bibitem{PhysRevLett.114.192502}
\bibinfo{author}{Kanungo, R.} \emph{et~al.}
\newblock \bibinfo{title}{Evidence of soft dipole resonance in
  $^{11}\mathrm{Li}$ with isoscalar character}.
\newblock \emph{\bibinfo{journal}{Phys. Rev. Lett.}}
  \textbf{\bibinfo{volume}{114}}, \bibinfo{pages}{192502}
  (\bibinfo{year}{2015}).

\bibitem{PhysRevC.80.064311}
\bibinfo{author}{Dussel, G.~G.}, \bibinfo{author}{Betan, R.~I.},
  \bibinfo{author}{Liotta, R.~J.} \& \bibinfo{author}{Vertse, T.}
\newblock \bibinfo{title}{Collective excitations in the continuum}.
\newblock \emph{\bibinfo{journal}{Phys. Rev. C}} \textbf{\bibinfo{volume}{80}},
  \bibinfo{pages}{064311} (\bibinfo{year}{2009}).

\bibitem{PhysRevC.51.38}
\bibinfo{author}{Wiringa, R.~B.}, \bibinfo{author}{Stoks, V. G.~J.} \&
  \bibinfo{author}{Schiavilla, R.}
\newblock \bibinfo{title}{Accurate nucleon-nucleon potential with
  charge-independence breaking}.
\newblock \emph{\bibinfo{journal}{Phys. Rev. C}} \textbf{\bibinfo{volume}{51}},
  \bibinfo{pages}{38--51} (\bibinfo{year}{1995}).

\bibitem{PhysRevC.63.024001}
\bibinfo{author}{Machleidt, R.}
\newblock \bibinfo{title}{High-precision, charge-dependent \text{B}onn
  nucleon-nucleon potential}.
\newblock \emph{\bibinfo{journal}{Phys. Rev. C}} \textbf{\bibinfo{volume}{63}},
  \bibinfo{pages}{024001} (\bibinfo{year}{2001}).

\bibitem{PhysRevC.68.041001}
\bibinfo{author}{Entem, D.~R.} \& \bibinfo{author}{Machleidt, R.}
\newblock \bibinfo{title}{Accurate charge-dependent nucleon-nucleon potential
  at fourth order of chiral perturbation theory}.
\newblock \emph{\bibinfo{journal}{Phys. Rev. C}} \textbf{\bibinfo{volume}{68}},
  \bibinfo{pages}{041001} (\bibinfo{year}{2003}).

\bibitem{Machleidt20111}
\bibinfo{author}{Machleidt, R.} \& \bibinfo{author}{Entem, D.~R.}
\newblock \bibinfo{title}{Chiral effective field theory and nuclear forces}.
\newblock \emph{\bibinfo{journal}{Physics Reports}}
  \textbf{\bibinfo{volume}{503}}, \bibinfo{pages}{1 -- 75}
  (\bibinfo{year}{2011}).

\bibitem{PhysRevD.48.5863}
\bibinfo{author}{G\l{}azek, S.~D.} \& \bibinfo{author}{Wilson, K.~G.}
\newblock \bibinfo{title}{Renormalization of hamiltonians}.
\newblock \emph{\bibinfo{journal}{Phys. Rev. D}} \textbf{\bibinfo{volume}{48}},
  \bibinfo{pages}{5863--5872} (\bibinfo{year}{1993}).

\bibitem{AnnPhys.506.77}
\bibinfo{author}{Wegner, F.}
\newblock \bibinfo{title}{Flow-equations for hamiltonians}.
\newblock \emph{\bibinfo{journal}{Ann. Phys. (Leipzig)}}
  \textbf{\bibinfo{volume}{506}}, \bibinfo{pages}{77--91}
  (\bibinfo{year}{1994}).

\bibitem{PhysRevLett.106.222502}
\bibinfo{author}{Tsukiyama, K.}, \bibinfo{author}{Bogner, S.~K.} \&
  \bibinfo{author}{Schwenk, A.}
\newblock \bibinfo{title}{In-medium similarity renormalization group for
  nuclei}.
\newblock \emph{\bibinfo{journal}{Phys. Rev. Lett.}}
  \textbf{\bibinfo{volume}{106}}, \bibinfo{pages}{222502}
  (\bibinfo{year}{2011}).

\bibitem{HERGERT2016165}
\bibinfo{author}{Hergert, H.}, \bibinfo{author}{Bogner, S.},
  \bibinfo{author}{Morris, T.}, \bibinfo{author}{Schwenk, A.} \&
  \bibinfo{author}{Tsukiyama, K.}
\newblock \bibinfo{title}{The in-medium similarity renormalization group: A
  novel ab initio method for nuclei}.
\newblock \emph{\bibinfo{journal}{Physics Reports}}
  \textbf{\bibinfo{volume}{621}}, \bibinfo{pages}{165 -- 222}
  (\bibinfo{year}{2016}).
\newblock \bibinfo{note}{Memorial Volume in Honor of Gerald E. Brown}.

\bibitem{PhysRevC.85.061304}
\bibinfo{author}{Tsukiyama, K.}, \bibinfo{author}{Bogner, S.~K.} \&
  \bibinfo{author}{Schwenk, A.}
\newblock \bibinfo{title}{In-medium similarity renormalization group for
  open-shell nuclei}.
\newblock \emph{\bibinfo{journal}{Phys. Rev. C}} \textbf{\bibinfo{volume}{85}},
  \bibinfo{pages}{061304} (\bibinfo{year}{2012}).

\bibitem{PhysRevLett.110.242501}
\bibinfo{author}{Hergert, H.}, \bibinfo{author}{Binder, S.},
  \bibinfo{author}{Calci, A.}, \bibinfo{author}{Langhammer, J.} \&
  \bibinfo{author}{Roth, R.}
\newblock \bibinfo{title}{Ab initio calculations of even oxygen isotopes with
  chiral two-plus-three-nucleon interactions}.
\newblock \emph{\bibinfo{journal}{Phys. Rev. Lett.}}
  \textbf{\bibinfo{volume}{110}}, \bibinfo{pages}{242501}
  (\bibinfo{year}{2013}).

\bibitem{PhysRevLett.118.032502}
\bibinfo{author}{Stroberg, S.~R.} \emph{et~al.}
\newblock \bibinfo{title}{Nucleus-dependent valence-space approach to nuclear
  structure}.
\newblock \emph{\bibinfo{journal}{Phys. Rev. Lett.}}
  \textbf{\bibinfo{volume}{118}}, \bibinfo{pages}{032502}
  (\bibinfo{year}{2017}).

\bibitem{PhysRevC.95.044304}
\bibinfo{author}{Parzuchowski, N.~M.}, \bibinfo{author}{Morris, T.~D.} \&
  \bibinfo{author}{Bogner, S.~K.}
\newblock \bibinfo{title}{Ab initio excited states from the in-medium
  similarity renormalization group}.
\newblock \emph{\bibinfo{journal}{Phys. Rev. C}} \textbf{\bibinfo{volume}{95}},
  \bibinfo{pages}{044304} (\bibinfo{year}{2017}).

\bibitem{PhysRevC.96.034324}
\bibinfo{author}{Parzuchowski, N.~M.}, \bibinfo{author}{Stroberg, S.~R.},
  \bibinfo{author}{Navr\'atil, P.}, \bibinfo{author}{Hergert, H.} \&
  \bibinfo{author}{Bogner, S.~K.}
\newblock \bibinfo{title}{Ab initio electromagnetic observables with the
  in-medium similarity renormalization group}.
\newblock \emph{\bibinfo{journal}{Phys. Rev. C}} \textbf{\bibinfo{volume}{96}},
  \bibinfo{pages}{034324} (\bibinfo{year}{2017}).

\bibitem{PhysRevLett.118.152503}
\bibinfo{author}{Gebrerufael, E.}, \bibinfo{author}{Vobig, K.},
  \bibinfo{author}{Hergert, H.} \& \bibinfo{author}{Roth, R.}
\newblock \bibinfo{title}{Ab initio description of open-shell nuclei: Merging
  no-core shell model and in-medium similarity renormalization group}.
\newblock \emph{\bibinfo{journal}{Phys. Rev. Lett.}}
  \textbf{\bibinfo{volume}{118}}, \bibinfo{pages}{152503}
  (\bibinfo{year}{2017}).

\bibitem{BERGGREN1968265}
\bibinfo{author}{Berggren, T.}
\newblock \bibinfo{title}{On the use of resonant states in eigenfunction
  expansions of scattering and reaction amplitudes}.
\newblock \emph{\bibinfo{journal}{Nuclear Physics A}}
  \textbf{\bibinfo{volume}{109}}, \bibinfo{pages}{265 -- 287}
  (\bibinfo{year}{1968}).

\bibitem{LIOTTA19961}
\bibinfo{author}{Liotta, R.}, \bibinfo{author}{Maglione, E.},
  \bibinfo{author}{Sandulescu, N.} \& \bibinfo{author}{Vertse, T.}
\newblock \bibinfo{title}{A representation to describe nuclear processes in the
  continuum}.
\newblock \emph{\bibinfo{journal}{Physics Letters B}}
  \textbf{\bibinfo{volume}{367}}, \bibinfo{pages}{1 -- 4}
  (\bibinfo{year}{1996}).

\bibitem{PhysRevLett.104.062701}
\bibinfo{author}{Tanaka, K.} \emph{et~al.}
\newblock \bibinfo{title}{Observation of a large reaction cross section in the
  drip-line nucleus $^{22}\mathbf{C}$}.
\newblock \emph{\bibinfo{journal}{Phys. Rev. Lett.}}
  \textbf{\bibinfo{volume}{104}}, \bibinfo{pages}{062701}
  (\bibinfo{year}{2010}).

\bibitem{PhysRevC.86.054604}
\bibinfo{author}{Kobayashi, N.} \emph{et~al.}
\newblock \bibinfo{title}{One- and two-neutron removal reactions from the most
  neutron-rich carbon isotopes}.
\newblock \emph{\bibinfo{journal}{Phys. Rev. C}} \textbf{\bibinfo{volume}{86}},
  \bibinfo{pages}{054604} (\bibinfo{year}{2012}).

\bibitem{PhysRevLett.109.202503}
\bibinfo{author}{Gaudefroy, L.} \emph{et~al.}
\newblock \bibinfo{title}{Direct mass measurements of $^{19}\mathrm{B}$,
  $^{22}\mathrm{C}$, $^{29}\mathrm{F}$, $^{31}\mathrm{Ne}$, $^{34}\mathrm{Na}$
  and other light exotic nuclei}.
\newblock \emph{\bibinfo{journal}{Phys. Rev. Lett.}}
  \textbf{\bibinfo{volume}{109}}, \bibinfo{pages}{202503}
  (\bibinfo{year}{2012}).

\bibitem{TOGANO2016412}
\bibinfo{author}{Togano, Y.} \emph{et~al.}
\newblock \bibinfo{title}{Interaction cross section study of the two-neutron
  halo nucleus $^{22}\mathrm{C}$}.
\newblock \emph{\bibinfo{journal}{Physics Letters B}}
  \textbf{\bibinfo{volume}{761}}, \bibinfo{pages}{412 -- 418}
  (\bibinfo{year}{2016}).

\bibitem{nndc}
\emph{\bibinfo{journal}{Data extracted using the NNDC On-Line Data Service,}}
  arXiv:\eprint{http://www.nndc.bnl.gov/}.

\bibitem{HOFFMAN200917}
\bibinfo{author}{Hoffman, C.~R.} \emph{et~al.}
\newblock \bibinfo{title}{Evidence for a doubly magic $^{24}\mathrm{O}$}.
\newblock \emph{\bibinfo{journal}{Physics Letters B}}
  \textbf{\bibinfo{volume}{672}}, \bibinfo{pages}{17 -- 21}
  (\bibinfo{year}{2009}).

\bibitem{PhysRevC.83.031303}
\bibinfo{author}{Hoffman, C.~R.} \emph{et~al.}
\newblock \bibinfo{title}{Observation of a two-neutron cascade from a resonance
  in $^{24}\mathrm{O}$}.
\newblock \emph{\bibinfo{journal}{Phys. Rev. C}} \textbf{\bibinfo{volume}{83}},
  \bibinfo{pages}{031303} (\bibinfo{year}{2011}).

\bibitem{PhysRevLett.110.192502}
\bibinfo{author}{Ekstr\"om, A.} \emph{et~al.}
\newblock \bibinfo{title}{Optimized chiral nucleon-nucleon interaction at
  next-to-next-to-leading order}.
\newblock \emph{\bibinfo{journal}{Phys. Rev. Lett.}}
  \textbf{\bibinfo{volume}{110}}, \bibinfo{pages}{192502}
  (\bibinfo{year}{2013}).

\bibitem{KUO199615}
\bibinfo{author}{Kuo, T.}, \bibinfo{author}{M\"uther, H.} \&
  \bibinfo{author}{Azimi-Nili, K.~A.}
\newblock \bibinfo{title}{Realistic effective interactions for halo nuclei}.
\newblock \emph{\bibinfo{journal}{Nuclear Physics A}}
  \textbf{\bibinfo{volume}{606}}, \bibinfo{pages}{15 -- 26}
  (\bibinfo{year}{1996}).

\bibitem{Sun:2018ekv}
\bibinfo{author}{Sun, X.~X.}, \bibinfo{author}{Zhao, J.} \&
  \bibinfo{author}{Zhou, S.~G.}
\newblock \bibinfo{title}{{Shrunk halo and quenched shell gap at $N=16$ in
  $^{22}$C: Inversion of $sd$ states and deformation effects}}.
\newblock \emph{\bibinfo{journal}{arXiv:1807.04991 [nucl-th]}}
  (\bibinfo{year}{2018}).

\bibitem{PhysRevLett.117.102501}
\bibinfo{author}{Kanungo, R.} \emph{et~al.}
\newblock \bibinfo{title}{Proton distribution radii of
  $^{12\char21{}19}\mathrm{C}$ illuminate features of neutron halos}.
\newblock \emph{\bibinfo{journal}{Phys. Rev. Lett.}}
  \textbf{\bibinfo{volume}{117}}, \bibinfo{pages}{102501}
  (\bibinfo{year}{2016}).

\bibitem{PhysRevC.81.064303}
\bibinfo{author}{Coraggio, L.}, \bibinfo{author}{Covello, A.},
  \bibinfo{author}{Gargano, A.} \& \bibinfo{author}{Itaco, N.}
\newblock \bibinfo{title}{Shell-model calculations for neutron-rich carbon
  isotopes with a chiral nucleon-nucleon potential}.
\newblock \emph{\bibinfo{journal}{Phys. Rev. C}} \textbf{\bibinfo{volume}{81}},
  \bibinfo{pages}{064303} (\bibinfo{year}{2010}).

\bibitem{PhysRevLett.113.142502}
\bibinfo{author}{Jansen, G.~R.}, \bibinfo{author}{Engel, J.},
  \bibinfo{author}{Hagen, G.}, \bibinfo{author}{Navratil, P.} \&
  \bibinfo{author}{Signoracci, A.}
\newblock \bibinfo{title}{Ab initio coupled-cluster effective interactions for
  the shell model: Application to neutron-rich oxygen and carbon isotopes}.
\newblock \emph{\bibinfo{journal}{Phys. Rev. Lett.}}
  \textbf{\bibinfo{volume}{113}}, \bibinfo{pages}{142502}
  (\bibinfo{year}{2014}).

\bibitem{1674-1137-41-10-104101}
\bibinfo{author}{Hu, B.~S.}, \bibinfo{author}{Wu, Q.} \& \bibinfo{author}{Xu,
  F.~R.}
\newblock \bibinfo{title}{{\it Ab initio} many-body perturbation theory and
  no-core shell model}.
\newblock \emph{\bibinfo{journal}{Chinese Physics C}}
  \textbf{\bibinfo{volume}{41}}, \bibinfo{pages}{104101}
  (\bibinfo{year}{2017}).

\bibitem{PhysRevC.94.014303}
\bibinfo{author}{Hu, B.~S.}, \bibinfo{author}{Xu, F.~R.}, \bibinfo{author}{Sun,
  Z.~H.}, \bibinfo{author}{Vary, J.~P.} \& \bibinfo{author}{Li, T.}
\newblock \bibinfo{title}{\textit{Ab initio} nuclear many-body perturbation
  calculations in the hartree-fock basis}.
\newblock \emph{\bibinfo{journal}{Phys. Rev. C}} \textbf{\bibinfo{volume}{94}},
  \bibinfo{pages}{014303} (\bibinfo{year}{2016}).

\bibitem{Morris}
\bibinfo{author}{Morris, T.~D.}
\newblock \emph{\bibinfo{title}{Systematic improvements of ab-initio in-medium
  similarity renormalization group calculations}} (\bibinfo{publisher}{Ph.D.
  thesis, Michigan State University, Michigan}, \bibinfo{year}{2016}).

\bibitem{jcp141}
\bibinfo{author}{Evangelista, F.~A.}
\newblock \bibinfo{title}{A driven similarity renormalization group approach to
  quantum many-body problems}.
\newblock \emph{\bibinfo{journal}{The Journal of Chemical Physics}}
  \textbf{\bibinfo{volume}{141}}, \bibinfo{pages}{054109}
  (\bibinfo{year}{2014}).

\bibitem{EVANGELISTA201227}
\bibinfo{author}{Evangelista, F.~A.} \& \bibinfo{author}{Gauss, J.}
\newblock \bibinfo{title}{On the approximation of the similarity-transformed
  hamiltonian in single-reference and multireference coupled cluster theory}.
\newblock \emph{\bibinfo{journal}{Chemical Physics}}
  \textbf{\bibinfo{volume}{401}}, \bibinfo{pages}{27 -- 35}
  (\bibinfo{year}{2012}).

\bibitem{PhysRevC.88.044318}
\bibinfo{author}{Papadimitriou, G.}, \bibinfo{author}{Rotureau, J.},
  \bibinfo{author}{Michel, N.}, \bibinfo{author}{P\l{}oszajczak, M.} \&
  \bibinfo{author}{Barrett, B.~R.}
\newblock \bibinfo{title}{Ab initio no-core gamow shell model calculations with
  realistic interactions}.
\newblock \emph{\bibinfo{journal}{Phys. Rev. C}} \textbf{\bibinfo{volume}{88}},
  \bibinfo{pages}{044318} (\bibinfo{year}{2013}).

\bibitem{0954-3899-16-2-013}
\bibinfo{author}{Rath, P.~K.}, \bibinfo{author}{Faessler, A.},
  \bibinfo{author}{M\"uther, H.} \& \bibinfo{author}{Watt, A.}
\newblock \bibinfo{title}{A practical solution to the problem of spurious
  states in shell-model calculations}.
\newblock \emph{\bibinfo{journal}{Journal of Physics G: Nuclear and Particle
  Physics}} \textbf{\bibinfo{volume}{16}}, \bibinfo{pages}{245}
  (\bibinfo{year}{1990}).

\bibitem{lawson}
\bibinfo{author}{Lawson, R.~D.}
\newblock \emph{\bibinfo{title}{Theory of the Nuclear Shell Model}}
  (\bibinfo{publisher}{Oxford University Press}, \bibinfo{year}{1980}).

\bibitem{PhysRevLett.103.062503}
\bibinfo{author}{Hagen, G.}, \bibinfo{author}{Papenbrock, T.} \&
  \bibinfo{author}{Dean, D.~J.}
\newblock \bibinfo{title}{Solution of the center-of-mass problem in nuclear
  structure calculations}.
\newblock \emph{\bibinfo{journal}{Phys. Rev. Lett.}}
  \textbf{\bibinfo{volume}{103}}, \bibinfo{pages}{062503}
  (\bibinfo{year}{2009}).

\bibitem{SUN2017227}
\bibinfo{author}{Sun, Z.~H.} \emph{et~al.}
\newblock \bibinfo{title}{Resonance and continuum gamow shell model with
  realistic nuclear forces}.
\newblock \emph{\bibinfo{journal}{Physics Letters B}}
  \textbf{\bibinfo{volume}{769}}, \bibinfo{pages}{227 -- 232}
  (\bibinfo{year}{2017}).

\bibitem{PhysRevC.96.044307}
\bibinfo{author}{Wang, S.~M.}, \bibinfo{author}{Michel, N.},
  \bibinfo{author}{Nazarewicz, W.} \& \bibinfo{author}{Xu, F.~R.}
\newblock \bibinfo{title}{Structure and decays of nuclear three-body systems:
  The gamow coupled-channel method in jacobi coordinates}.
\newblock \emph{\bibinfo{journal}{Phys. Rev. C}} \textbf{\bibinfo{volume}{96}},
  \bibinfo{pages}{044307} (\bibinfo{year}{2017}).

\bibitem{PhysRevC.73.064307}
\bibinfo{author}{Hagen, G.}, \bibinfo{author}{Hjorth-Jensen, M.} \&
  \bibinfo{author}{Michel, N.}
\newblock \bibinfo{title}{Gamow shell model and realistic nucleon-nucleon
  interactions}.
\newblock \emph{\bibinfo{journal}{Phys. Rev. C}} \textbf{\bibinfo{volume}{73}},
  \bibinfo{pages}{064307} (\bibinfo{year}{2006}).

\bibitem{PhysRevC.92.034331}
\bibinfo{author}{Morris, T.~D.}, \bibinfo{author}{Parzuchowski, N.~M.} \&
  \bibinfo{author}{Bogner, S.~K.}
\newblock \bibinfo{title}{Magnus expansion and in-medium similarity
  renormalization group}.
\newblock \emph{\bibinfo{journal}{Phys. Rev. C}} \textbf{\bibinfo{volume}{92}},
  \bibinfo{pages}{034331} (\bibinfo{year}{2015}).

\bibitem{ragnar}
\bibinfo{author}{Stroberg, S.~R.}
  arXiv:\eprint{https://github.com/ragnarstroberg/ragnar\_imsrg}.

\end{thebibliography}

\section*{Acknowledgements}

Valuable discussions with Simin Wang, N. Michel, M. P\l{}oszajczak, G. Hagen
and C.W. Johnson are gratefully acknowledged.
The valence-space IMSRG code used is Ragnar$\_$IMSRG \cite{ragnar}.
This work has been supported by
the National Key R${\&}$D Program of China under Grant No. 2018YFA0404401;
the National Natural Science Foundation of China under Grants No. 11835001, No. 11320101004 and No. 11575007;
China Postdoctoral Science Foundation under Grant No. 2018M630018;
and the CUSTIPEN (China-U.S. Theory Institute for Physics with Exotic Nuclei) funded by the U.S.  Department of Energy,
Office of Science under Grant No. DE-SC0009971.
We acknowledge the High-performance Computing Platform of Peking University for providing computational resources.

\section*{Electronic supplementary material}

\begin{figure}
\centering
\includegraphics[scale=0.7]{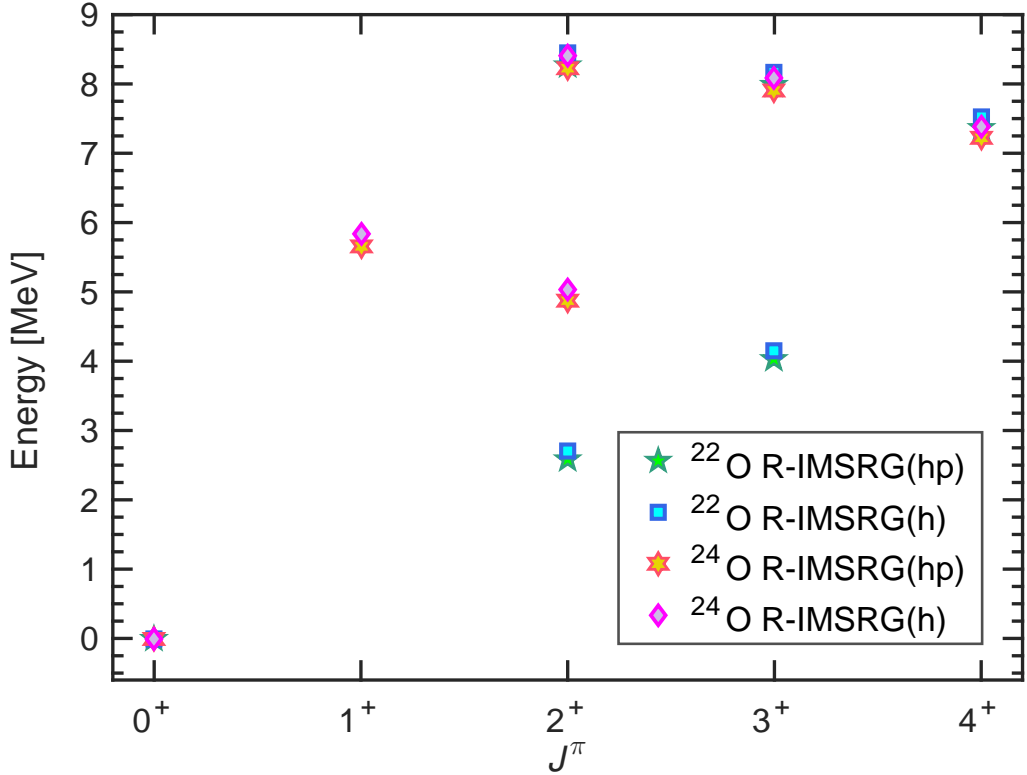}
\caption{Excited-state energies in $^{22,24}$O, calculated by different MAGNUS(2*) approximations. The R-IMSRG(hp) means the full hole-hole and particle-particle excitations are included in the auxiliary one-body operator of the real-energy MAGNUS(2*) approach, while R-IMSRG(h) indicates only the hole-hole excitations are considered.}
\label{goosetank}
\end{figure}

In the MAGNUS(2*) approach, an auxiliary one-body operator with hole-hole and particle-particle excitations is introduced \cite{Morris,jcp141,PhysRevC.95.044304}.
It has been shown~\cite{EVANGELISTA201227} that the contribution of particle-particle excitations is significantly smaller than the hole-hole contribution in chemistry systems.
Figure \ref{goosetank} compares the calculations with only the hole-hole excitations and the calculation with full hole-hole and particle-particle excitations in $^{22,24}$O.
We see that the approximation with only the hole-hole excitations gives almost the same results as the full MAGNUS(2*) approach, implying that the contribution from particle-particle excitations can be neglected. 




\end{document}